\documentclass[12pt]{article}
\usepackage{putex}
\usepackage{latexsym}

\newcommand{\beq}{\begin{eqnarray}}
\newcommand{\eeq}{\end{eqnarray}}

\newcommand{\bi}{\bibitem}

\newcommand{\be}{\begin{equation}}
\newcommand{\ee}{\end{equation}}
\newcommand{\ben}{\begin{eqnarray}\displaystyle}
\newcommand{\een}{\end{eqnarray}}

\begin{document}

\preprint{hep-th/0405282 \\ PUPT-2120 \\ NSF-KITP-04-71}

\institution{PU}{Joseph Henry Laboratories,\cr Princeton University, Princeton, NJ 08544}
\institution{KITP}{Kavli Institute for Theoretical Physics,\cr University of California, Santa Barbara, CA  93106}

\title{Symmetry Breaking and Axionic Strings
in the Warped Deformed Conifold}

\authors{Steven S.~Gubser,\worksat{\PU,}\footnote{e-mail: \tt 
ssgubser@Princeton.EDU} 
Christopher P.~Herzog,\worksat{\KITP,}\footnote{e-mail: \tt 
herzog@kitp.ucsb.EDU} 
and Igor R.~Klebanov\worksat{\PU,}\footnote{e-mail: \tt klebanov@Princeton.EDU}}

\abstract{
We interpret D-strings at the bottom of the warped deformed
conifold as axionic strings in the dual cascading 
$SU(N+M)\times SU(N)$ gauge theory. The axion is a massless pseudo-scalar
glueball which we find in the supergravity fluctuation spectrum
and interpret
as the Goldstone boson of spontaneously broken $U(1)$
baryon number symmetry. The existence of this
massless glueball, anticipated in hep-th/0101013, 
supports the idea
that the cascading gauge theory is on the baryonic branch,
i.e. the $U(1)_B$ global symmetry is broken by expectation values of
baryonic operators. We also find a massless scalar glueball,
which is a superpartner of the pseudo-scalar. This scalar mode is a 
mixture of an NS-NS 2-form and a
metric perturbation of the warped deformed conifold of a type first
considered in hep-th/0012034.}

\maketitle

\section{Introduction}

One of the goals of the string formulation of gauge theories is
to find a dual string description of confinement \cite{Polyakov}.
Indeed, a dual approach
to conformal theories, the AdS/CFT correspondence 
\cite{Maldacena,Gubser,Witten:1998qj},
extends to confining backgrounds.
Witten \cite{Witten:1998zw} initiated 
such an approach to non-supersymmetric theories, 
while the authors of \cite{KS,MN} found 
supergravity solutions dual to ${\cal N}=1$
supersymmetric confining theories.
However, the UV structure of these confining theories does not
correspond to the conventional asymptotically free gauge theory.
In \cite{Witten:1998zw,MN} the UV theory exists in a dimension
higher than 4 and is not field-theoretic.
In \cite{KS} the gauge theory remains 4-dimensional
but has $SU(N+M) \times SU(N)$ gauge group \cite{GK} whose flow exhibits an RG
cascade \cite{KS,KT}. The fact that the gauge theory remains 4-dimensional in
the UV allows for comparison of the beta-function coefficients in
this case \cite{Herzog}, but neither gauge group is asymptotically free.

While the UV physics of the confining backgrounds in \cite{KS,MN}
is exotic, it was generally hoped that they are in the same IR universality
class as the 
pure glue ${\cal N}=1$ gauge theory. In this paper we study this question
in some detail for the warped deformed conifold
of \cite{KS}, and reach a conclusion that this is
not quite the case.  An old puzzle guides our investigation: what is the
gauge theory interpretation of D1-branes in the deformed conifold
background \cite{KS}? 
The interpretation of the fundamental strings
placed in the IR region of the metric is clear:
they are dual to confining strings,
and this duality can be used to extract results 
on $k$-string tensions \cite{Herzog:2001fq} (see also \cite{Hartnoll}).
Like the fundamental strings, the D-strings fall to the bottom of
the throat, $\tau=0$, where they
remain tensionful; hence, they cannot be dual to `t Hooft loops which must
be screened \cite{KS}.

Two options remain for resolving this puzzle. The first one is that
there is an effect that destabilizes the D-strings or removes them from
the IR region. For example, D-strings could blow up into
a D3-brane wrapped over an $S^2$ at some finite $\tau$, similar
to how F-strings blow up into a D3-brane wrapping an
$S^2$ at some finite azimuthal angle of the $S^3$. An implication
of the latter effect is that a composite state of $M$ F-strings does
not have a tension, in agreement with the dual gauge theory 
\cite{Herzog:2001fq}.
A detailed study of this effect, presented in Appendix B, shows
that this blow-up does not occur for D-strings.
The fact that
we have not found any destabilizing effect or blow-up of D-strings
 leads us to believe that they are stable objects.
Therefore, we have to explore this other option and look
at their dual interpretation in the 
gauge theory. 

We propose that in the dual gauge theory they are global strings
that create a monodromy of a massless axion field.\footnote{
We are grateful to
E. Witten for emphasizing this possibility to us.}
For this explanation to make sense, the IR gauge theory must differ from the
pure glue ${\cal N}=1$ theory in that it contains a massless 
pseudo-scalar bound state (glueball)
that plays the role of the axion field.
The study of glueball spectra in the background of \cite{KS} is complicated
due to mixing of various perturbations. Glueball spectra corresponding to
a small subset of fluctuations were calculated in 
\cite{Krasnitz,Caceres,Amador}, and no massless modes were found. 
The fact that the massless mode must couple directly to a D-string means that
it corresponds to a certain perturbation of the RR 2-form field
which also mixes with the RR 4-form field.
We turn to the necessary ansatz in sections 2 and 3,
and indeed find a massless glueball. This mode should
be interpreted as the Goldstone boson of spontaneously
broken global $U(1)$ baryon number symmetry. 
This Goldstone mode was anticipated by 
Aharony in \cite{Aharony},\footnote{I.R.K. acknowledges 
very useful discussions on this issue
with O. Aharony and M. Strassler early in 2001. } and
its presence supports the claim made in \cite{KS,Aharony} that
the cascading gauge theory is on the baryonic branch \cite{APS}, i.e. certain
baryonic operators acquire expectation values. This interpretation
of the massless glueball is presented in section 4.
The supersymmetric Goldstone mechanism gives rise also to a massless scalar
mode or ``saxion.'' In section 5 the supergravity dual of this mode is identified
as a massless glueball coming from a mixture of an NS-NS 2-form and
a metric deformation. The ansatz for such perturbations
was written down some time ago by Papadopoulos and Tseytlin \cite{PT}.

Besides being an interesting example of the gauge/gravity duality,
the warped deformed conifold background offers 
interesting possibilities for solving the hierarchy problem \cite{KS,GKP}.
If the background is embedded into a compact CY space with NS-NS and R-R
fluxes, then an exponential
hierarchy may be created between the UV compactification scale and
the IR scale at the bottom of the throat. Models of this type received
an additional boost due to a possibility of fixing all moduli proposed in
\cite{KKLT}, and a subsequent exploration of cosmology in
\cite{KKLMMT}. In interesting recent papers \cite{Copeland}, 
a new role was proposed
for various $(p,q)$ strings placed in the IR region. Besides being
the confining or axionic strings from the point of view of the gauge theory, 
they
may be realizations of cosmic strings. The exponential warping of the
background lowers the tension significantly, and makes them 
plausible cosmic string candidates. Our improved understanding of D-strings 
in the warped deformed conifold 
as axionic strings in the gauge theory
should therefore have new implications for
this cosmological modeling.
In section 5 we discuss the Higgs mechanism that occurs
upon embedding the warped deformed conifold into a flux compactification,
and make a connection with recent papers on 
Abrikosov-Nielsen-Olesen strings in supergravity \cite{Dvali} 
(see also earlier work by \cite{Edelstein}).
In section 7 we find the supergravity background describing
D-strings at $\tau=0$ that are smeared over their transverse directions
$x^2, x^3$ as well as over the $S^3$. This
may be thought of as a non-commutative deformation of the
warped deformed conifold. 
We conclude in section 8.

\section{Ansatz for the 2-form perturbation}

As a warmup, we first look for our massless
pseudo-scalar glueball in the UV asymptotic form of the 
warped conifold background constructed in \cite{KT}. The metric and hence the calculation are simpler than in the full solution of \cite{KS}, but the qualitative features of our ansatz will be the same.

To begin, consider a D1-brane extended in two of the four dimensions
in ${\bf R}^{3,1}$.
Because the D1-brane carries electric charge under the R-R three-form $F_3$, it is natural to think that a pseudoscalar
$a$ in four dimensions, defined so that $*_4 da = \delta F_3$, experiences monodromy as one loops around the D1-brane world-volume.  
But it is 
complicated to write down a full ansatz for the field generated by a D1-brane.  
A simpler ansatz is $a(t) \sim t$, which is to say 
$\delta F_3 \sim f_1 dx^1 \wedge dx^2 \wedge dx^3$.  
The idea is to add whatever else we need to make this a solution to the 
linearized equations of motion.  A solution so obtained would 
represent a zero-momentum pseudoscalar.  In section~\ref{GENERALIZE} 
we will extend our treatment to arbitrary light-like momenta, which is 
to say arbitrary harmonic $a(t,x^1,x^2,x^3)$.

Using the conventions of \cite{Herzog}, 
the equations of motion include
 \eqn{eoms}{
  d (e^{-\phi} * H_3) = - g_s^2 F_5 \wedge F_3 \,,\quad 
d( e^\phi *F_3) = F_5 \wedge H_3 \,,\quad
  F_5 = * F_5 \,,
 }
while the Bianchi identities include
 \eqn{bianchis}{
  dF_3 = 0 \,,\quad dH_3 = 0 \,,\quad dF_5 = H_3 \wedge F_3 \,.
 }
The background solution \cite{KT} has
 \eqn{Background}{
  ds^2 &= {1 \over \sqrt{h}} (-dt^2 + d\vec{x}^2) + 
   \sqrt{h} (dr^2 + r^2 ds_{T^{11}}^2)\ ,  \cr
  F_3 &= {M\alpha' \over 2} \omega_3\ , \quad
  H_3 = {3 g_s M\alpha' \over 2r} dr \wedge \omega_2 \ ,\quad
  B_2 = {3 g_s M\alpha' \over 2} \log {r \over r_*} \; \omega_2\ ,  \cr
  F_5 &= (1 + *) B_2 \wedge F_3 \,.
 }
where $\omega_2$ and $\omega_3$ are defined in Appendix A, (\ref{defo2}) and (\ref{defo3}),
$T^{1,1}$ is the level surface of the conifold, and $h(r)$ is a harmonic function on the conifold.

Hodge duals can be defined through the equations
 \eqn{HodgeDef}{
  \nu \wedge *\nu = \nu^2 \vol \ ,\qquad
   \nu^2 = {1 \over p!} g^{i_1j_1} \cdots g^{i_pj_p}
    \nu_{i_1 \cdots i_p} \nu_{j_1 \cdots j_p} \,,
 }
and for the background solution,
 \eqn{Bvol}{
  \vol = {\sqrt{h} r^5 \over 54} dt \wedge dx^1 \wedge dx^2 \wedge
   dx^3 \wedge \omega_2 \wedge \omega_3 \wedge dr \,.
 }
Some important identities in the following calculations are
 \eqn{Useful}{
  \omega_3 = g^5 \wedge \omega_2 \ ,\quad
  dg^5 \wedge \omega_2 = 0 \ ,\quad
  dg^5 \wedge dg^5 = -4 \omega_2 \wedge \omega_2 \,,
 }
where $g^5$ is also defined in Appendix A.

The perturbation ansatz we adopt is
 \eqn{Perturbation}{
  \delta H_3 &= 0 \ , \cr
  \delta F_3 &= f_1 dx^1 \wedge dx^2 \wedge dx^3 + 
   f_2 dt \wedge dg^5 + f_2' dt \wedge dr \wedge g^5\ ,  \cr
  \delta F_5 &= f_1 dx^1 \wedge dx^2 \wedge dx^3 \wedge B_2 +
   f_1 g_s h r \log r \; dt \wedge dr \wedge F_3 \,.
 }
The variations of all other fields, including the metric and the dilaton, vanish.
The Bianchi identity for $F_3$ implies that $f_1$ is a constant.  
The second two terms in $\delta F_3$ sum to the exact form $-d (f_2 dt \wedge g^5)$.
It's easy to verify that 
 \eqn{TwoLinearizedEOMS}{
  d\delta F_5 &= H_3 \wedge \delta F_3\ ,  \cr
  g_s^{-2} d*\delta H_3 & = -\delta F_5 \wedge F_3 - F_5 \wedge \delta F_3
   \cr &= -f_1 dx^1 \wedge dx^2 \wedge dx^3 \wedge B_2 \wedge F_3 - 
    B_2 \wedge F_3 \wedge f_1 dx^1 \wedge dx^2 \wedge dx^3=0 \,.
 }
The remaining linearized equation of motion is
 \eqn{LastEOM}{
  d*\delta F_3 = \delta F_5 \wedge H_3 \,.
 }
To show that this can be satisfied, note first that
 \eqn{HodgeDeltaF}{
  *\delta F_3 &= -f_1 {h^2 r^5 \over 54} dt \wedge \omega_2 \wedge
    \omega_3 \wedge dr - 
   f_2 {r \over 3} dx^1 \wedge dx^2 \wedge dx^3 \wedge dr \wedge
    g^5 \wedge dg^5   \cr &\qquad{} + 
   f_2' {r^3 \over 6} dx^1 \wedge dx^2 \wedge dx^3 \wedge \omega_2
    \wedge \omega_2  \ ,\cr
  \delta F_5 \wedge H_3 &= f_1 dx^1 \wedge dx^2 \wedge dx^3 \wedge
   B_2 \wedge H_3 \,.
 }
Evidently, the left and right hand sides of \LastEOM\ are 
both proportional to $dx^1 \wedge dx^2 \wedge dx^3 \wedge dr \wedge \omega_2 \wedge \omega_2$.
Collecting terms, we find
 \eqn{fDiffEQ}{
  \left[- \partial_r {r^3 \over 6} \partial_r + {4r \over 3} \right]
   f_2 = \left( {3 g_s M\alpha' \over 2} \right)^2
    {\log r/r_* \over r} f_1 \,,
 }
and we recall that $f_1$ is a constant. 
In principle, \fDiffEQ\ may be solved for $f_2$, but because the background is singular 
at finite $r$, the solution doesn't mean very much: 
we need the full deformed conifold solution in order to set boundary conditions in 
the IR before we integrate out to the UV region.

\section{Generalization to the Warped Deformed Conifold}
\label{GENERALIZE}

It turns out that there is a simple generalization of the 
ansatz (\ref{Perturbation}) that is consistent for the complete KS background
\cite{KS}.  Relevant formulae for this background are collected in Appendix A.  The ansatz is
\eqn{NewPerturbation}{
  \delta H_3 &= 0 \ , \cr
  \delta F_3 &= f_1 *_4 da + 
   f_2(\tau) da \wedge dg^5 + f_2' da \wedge d\tau \wedge g^5\ ,  \cr
  \delta F_5 &= (1 + *) \delta F_3 \wedge B_2=
f_1 (*_4 da
- {\epsilon^{4/3}\over 6 K^2(\tau)} h(\tau) da\wedge d\tau\wedge g^5) \wedge B_2 \,,
 }
where now $f_2'= d f_2/d\tau$, $h(\tau)$ is
given by (\ref{intsol}), and $K(\tau)$ by (\ref{Keqn}).  
We have generalized from $a \sim t$ to an arbitrary harmonic 
function $a(t,x^1,x^2,x^3)$: thus $d*_4 da = 0$.\footnote{
The 4-dimensional Hodge dual $*_4$ is calculated with the Minkowski metric,
${\rm vol}_4= dt\wedge dx^1\wedge dx^2\wedge dx^3$.}
  As before, the second two terms in 
$\delta F_3$ sum to the exact form $-d (f_2 da \wedge g^5)$. 
The equation $d\delta F_3=0$ requires that $f_1$ is constant; we will set
$f_1=1$ from here on.
The equation $d\delta F_5 = H_3 \wedge \delta F_3$ is satisfied for harmonic $a$ due 
to the identities
\begin{equation}
H_3\wedge d\tau \wedge g^5=0\ , \qquad
H_3\wedge dg^5=H_3\wedge (g^1\wedge g^4+g^3\wedge g^2)=0
\,.
\end{equation}

The Hodge duals are computed with the help of
\begin{equation}
\vol = {\epsilon^4\over 96} h^{1/2} \sinh^2\tau
dt \wedge dx^1 \wedge dx^2 \wedge dx^3\wedge d\tau \wedge g^1\wedge g^2
\wedge g^3\wedge g^4\wedge g^5 \,.
\end{equation}
Using the explicit form of $B_2$,
\begin{equation}
B_2=  {g_s M\alpha' (\tau\coth\tau -1)\over 2\sinh \tau}
\left [\sinh^2 \left ({\tau\over 2}\right ) g^1\wedge g^2+
\cosh^2 \left ({\tau\over 2}\right ) g^3\wedge g^4 \right ] \ ,
\end{equation} 
we can check that the second term in $\delta F_5$ is Hodge dual to the first.
Acting with an exterior derivative on this second term makes it vanish.

We also have
 \eqn{NewHodgeDeltaF}{
*\delta F_3 &= h^2 \sinh^2 \tau
{\epsilon^{4/3}\over 96} da\wedge d\tau \wedge g^1\wedge g^2
\wedge g^3\wedge g^4\wedge g^5 \cr
 &\qquad{} -f_2 {\epsilon^{4/3}\over 6 K^2} 
(*_4 da) \wedge d\tau \wedge g^5\wedge dg^5\cr
 &\qquad{} +f_2' {3\epsilon^{4/3}K^4 \sinh^2 \tau \over 8}
 (*_4 da) \wedge g^1\wedge g^2
\wedge g^3\wedge g^4
}
and
\begin{equation}
\delta F_5\wedge H_3=  {(g_s M\alpha')^2\over 4} (f(\tau) k(\tau))' 
(*_4 da) \wedge d\tau \wedge g^1\wedge g^2
\wedge g^3\wedge g^4\ ,
\end{equation}
where $f(\tau) k(\tau) =(\tau\coth \tau -1)^2/4 $.

The second equation in (\ref{TwoLinearizedEOMS}) may be checked
exactly as in the preceding section.\footnote{ Note that the ability to satisfy
this equation with $\delta H_3=0$ depends crucially on having
the background values of $B_2$ and $F_3$, in addition to $F_5$.
There are no similar solutions for backgrounds where only $F_5$ is present.}
 From the remaining equation (\ref{LastEOM}), for harmonic $a$, we then get
\begin{equation}\label{threef}
-{d\over d\tau} [K^4 \sinh^2 \tau f_2'] + {8\over 9 K^2} f_2 =
{ (g_s M \alpha')^2\over 3 \epsilon^{4/3} } (\tau\coth \tau -1)\left (\coth\tau -
{\tau\over \sinh^2\tau}\right ) \,.
\end{equation}

\subsection{Solving the Differential Equation}
\label{SOLVING}

To solve this equation, let us first note that the homogeneous equation
\begin{equation}
-{d\over d\tau} [K^4 \sinh^2 \tau y'] + {8\over 9 K^2} y =0
\end{equation}
has the two solutions 
\begin{equation}
y_1(\tau)= [\sinh(2\tau)- 2\tau]^{1/3}\ ,
\qquad
y_2(\tau)= [\sinh(2\tau)- 2\tau]^{-2/3}
\ .\end{equation}
 From the theory of second-order ODE's it is known that
the general solution of inhomogeneous equations of the form
\begin{equation}
y'' + p(\tau) y' + q(\tau) y = g(\tau)
\end{equation}
is
\begin{equation}
y = c_1 y_1(\tau) + c_2 y_2(\tau) + Y(\tau)\ .
\end{equation}
A particular solution is
\begin{equation}
Y(\tau) = -y_1(\tau) \int_0^\tau y_2(x) {g(x) \over W(y_1, y_2)(x)} dx 
+y_2(\tau) \int_0^\tau y_1(x) {g(x) \over W(y_1, y_2)(x)} dx\ ,  
\end{equation}
where the Wronskian
\begin{equation}
W(y_1, y_2) = y_1 y_2' - y_2 y_1' \ .
\end{equation}

Using this method we find that the solution of (\ref{threef})
that is regular both for small
and for large $\tau$ is
 \eqn{FoundFTwo}{
  f_2 (\tau) &= c \left[ -K h(\tau) \sinh \tau + 
   {1 \over K^2 \sinh^2 \tau} \int_0^\tau dx \, h' K^3 \sinh^3 x  
    \right] \cr &=
   -{2 c \over K^2 \sinh^2 \tau}\int_0^\tau dx \, h(x) \sinh^2 x \,,
 }
where $c \sim \epsilon^{4/3}$.
This implies that
$f_2\sim \tau$ for small $\tau$, and $f_2\sim \tau e^{-2\tau/3}$ for large $\tau$.

We check that $\delta F_3$ is normalizable by integrating
$\sqrt{-G} |\delta F_3|^2$ over $\tau$ where $G$ is the ten dimensional
metric.
Note that
\be
\sqrt{-G} |\delta F_3|^2 d^{10}x = \delta F_3 \wedge * \delta F_3 \ .
\ee
 From (\ref{NewPerturbation}) and (\ref{NewHodgeDeltaF}), we see that
we need to check the convergence of three separate $\tau$ integrals.
These integrals are
\begin{eqnarray}
 \int_0^\infty f_1^2 h(\tau)^2 \sinh(\tau)^2 d\tau \ , \nonumber \\
 \int_0^\infty \frac{f_2(\tau)^2}{K^2} d\tau \ , \nonumber \\
 \int_0^\infty (f_2')^2 K^4 \sinh(\tau)^2 d\tau \ .
\end{eqnarray}

Each of these three integrals is well behaved at small $\tau$ and falls
off as $e^{-2\tau/3}$ at large $\tau$.  
So, this is indeed a normalizable zero-mode of the KS background.

It is interesting to compare this calculation with a similar calculation for
the RR-charged MN background \cite{MN},
\begin{equation}
ds_{str}^2 = e^\phi \left [ dx_{||}^2 +
N(d\rho^2 + e^{2g} d\Omega^2) + {1\over 4}\sum_{a=1}^3 (\omega^a- A^a)^2
\right ]
\ ,
\end{equation}
where 
\begin{equation}
e^{2\phi(\rho)}= g_s^2 {\sinh (2\rho)\over 2 e^{g(\rho)}}
\ ,\qquad e^{2g(\rho)}= \rho \coth(2\rho)- {\rho^2\over \sinh^2 (2\rho)}-
{1\over 4}\ .
\end{equation}
In this case it is not hard to check that $\delta F_3= *_4 da$
is a solution of the equations of motion, with all other variations set to zero.
However, for the MN metric, the norm
\be \label{normMN}
\int \sqrt{-G} |\delta F_3|^2 d^{10}x \sim \int_0^\infty d\rho e^{2\phi+2g}
\ee
has an exponential divergence at large $\rho$. 
The large $\rho$ limit of the MN
background is the metric of D5-branes, which is a linear dilaton background.
Such a UV boundary condition generally results in a continuous spectrum of glueballs
above a certain gap, i.e. their wavefunctions are not normalizable \cite{Aha,Amet},
with a linear divergence of the norm. The fact that the divergence in
(\ref{normMN}) is exponential suggests that this mode should be
excluded.\footnote{We are grateful to the referee for this
suggestion.}
Indeed, in the Maldacena-Nunez case there is no obvious
continuous symmetry breaking that could produce a Goldstone mode.

\section{Interpretation of the Zero-Mode}
\label{INTERPRETATION}

Clearly, the zero-mode we have found is a rather special phenomenon
that is tied to the properties of the KS background. Usually,
zero-mass particles exist due to some spontaneously 
broken symmetry. We would like to argue, following \cite{Aharony},
that the zero-mass glueball we are finding is due to the spontaneously
broken global $U(1)$ baryon number symmetry.

Recall that in the context of AdS/CFT correspondence, global symmetries
in the boundary CFT become gauge symmetries in the dual bulk supergravity
description.  Consider for a moment the simpler parent of the KS background,
i.e. $AdS_5 \times T^{1,1}$ dual to a superconformal $SU(N) \times SU(N)$
gauge theory \cite{KWit,MP}. 
In this simpler theory, the $U(1)$ baryon number symmetry
is preserved, and the gauge field $A$ dual to the baryon number current
$J^\mu$ is identified as $\delta C_4\sim \omega_3\wedge A$ \cite{KW,Ceres}. 
This CFT also possesses a global $U(1)_R$ symmetry, as well as $SU(2)\times
SU(2)$.

In the KS background the $SU(2)\times SU(2)$
is preserved, but the $U(1)_R$ symmetry is broken
in the UV by the chiral anomaly down to ${\bf Z}_{2M}$ \cite{KOW}.
Further spontaneous breaking of this discrete
symmetry to ${\bf Z}_2$ does not lead to the appearance of
a Goldstone mode. The $U(1)_B$ symmetry is not anomalous,
and its spontaneous breaking does lead to the appearance
of a Goldstone mode which we found above.
 The form of the
$\delta F_5$ in (\ref{NewPerturbation}) makes the connection between
our zero-mode and $U(1)_B$ evident.
Asymptotically, at large $\tau$, 
there is a component $\sim \omega_3\wedge da\wedge d\tau$ in $\delta F_5$.
Thus, in our case, we have 
$A\sim da$. For the 4-d effective Lagrangian,
there should be a coupling 
\begin{equation} \label{Jda}
{1\over f_a}\int d^4 x J^\mu \partial_\mu a 
=-{1\over f_a}\int d^4 x \, a(x) (\partial_\mu J^\mu) \ ,
\end{equation}
i.e. the Goldstone boson $a$ enters as the parameter of the baryon number
transformation. 
 It is important that this transformation
does not leave the vacuum invariant!
As discussed in \cite{KS,Aharony}
the theory is on the baryonic branch: for example,
``the last step'' of the cascade takes place through
giving expectation values to baryonic operators in the
$SU(2M)\times SU(M)$ gauge theory
coupled to bifundamental fields
$A_i, B_j$, $i,j=1,2$. 
From the point of view of the $SU(2M)$, the number of flavors equals the
number of colors. Hence, 
in addition to mesonic operators
$(N_{ij})^\alpha_\beta$, the gauge theory has 
baryonic operators invariant under the $SU(2M)\times SU(M)$
gauge symmetry:
\eqn{baryonops}{
{\cal B}& \sim \epsilon_{\alpha_1\alpha_2\ldots \alpha_{2M}}
(A_1)^{\alpha_1}_1 (A_1)^{\alpha_2}_{2}\ldots
(A_1)^{\alpha_M}_{M} 
(A_2)^{\alpha_{M+1} }_{1} (A_2)^{\alpha_{M+2} }_{2}\ldots
(A_2)^{\alpha_{2M} }_{M} \ ,\cr
\bar {\cal B}& \sim \epsilon^{\alpha_1\alpha_2\ldots
\alpha_{2M}}
(B_1)^{1}_{\alpha_1} (B_1)^{2}_{\alpha_2}\ldots
(B_1)^{M}_{\alpha_M} 
(B_2)^{1}_{\alpha_{M+1} } 
(B_2)^{2}_{\alpha_{M+2} }\ldots 
(B_2)^{M}_{\alpha_{2M} }\ . 
}
The baryonic operators are also invariant under
the $SU(2)\times SU(2)$ global symmetry rotating $A_i,B_j$.
The superpotential has the form
\begin{equation}
W = \lambda (N_{ij})^\alpha_\beta
(N_{k\ell})_\alpha^\beta\epsilon^{ik}\epsilon^{j\ell}
+ X(\det [(N_{ij})^\alpha_\beta]
-{\cal B}\bar{\cal B} - \Lambda_{2M}^{4M}) \ .
\end{equation}

The supersymmetry-preserving vacua include the baryonic branch:
\eqn{superW}{
X = 0 \ ; \ N = 0 \ ;\quad 
 {\cal B} \bar{\cal B} = -\Lambda_{2M}^{4M} \ ,
}
where the $SO(4)$ global symmetry 
rotating $A_i,B_j$
is unbroken.  Since the supergravity background of \cite{KS}
also has this symmetry, it is natural to identify
the dual of this background with the baryonic branch of
the cascading theory. 
The expectation values of the baryonic operators
spontaneously break the $U(1)$ baryon number symmetry
$A_k \to e^{i\alpha} A_k$, $B_j\to e^{-i\alpha} B_j$.
The deformed conifold as described in \cite{KS} 
corresponds to a vacuum where 
$|{\cal B}| = |\bar{\cal B}|=\Lambda_{2M}^{2M}$, which is
invariant under the exchange of the
$A$'s with the $B$'s accompanied by charge conjugation
in both gauge groups. As noted in \cite{Aharony}, 
the baryonic branch has complex dimension $1$, 
and it can be parametrized by $\xi$ where
 \eqn{xiDef}{
  {\cal B} = i\xi \Lambda_{2M}^{2M} \,,\qquad
  \bar{\cal B} = {i \over \xi} \Lambda_{2M}^{2M} \,.
 }
The pseudo-scalar Goldstone mode must correspond to 
changing $\xi$ by a phase, 
since this is precisely what a $U(1)_B$ symmetry transformation does.  
As usual, the gradient of
the pseudo-scalar Goldstone mode 
$f_a\partial_\mu a$ is created from the vacuum by the action of 
the axial baryon number current, $J_\mu$ (we expect that the scale
of the dimensionful `decay constant' $f_a$ is determined by the
baryon expectation values).
To summarize, we see that the breaking of the $U(1)$ baryon number symmetry
necessitates the presence of a massless 
pseudo-scalar glueball, which we have found.

By supersymmetry, our pseudoscalar Goldstone mode
falls into a massless ${\mathcal N}=1$
chiral multiplet.  Hence, there will also be a massless 
scalar mode or ``saxion''
and corresponding Weyl fermion or ``axino.''  The saxion must
correspond to changing $\xi$ by a positive real factor.  

Let us examine more fully the superfield which contains this 
$U(1)$ baryon number current: it is
 \eqn{Udef}{
  {\cal U} = \tr A_i \bar{A}_i - \tr B_j \bar{B}_j \,,
 }
and $J^\mu \sigma^\mu_{\alpha\dot\alpha}$ appears as the 
$\theta^\alpha \bar\theta^{\dot\alpha}$ 
term.\footnote{For simplicity, we have suppressed the
 $e^{V}$ factors in \Udef\ which are necessary for ${\cal U}$ to be 
a gauge-invariant operator.} The lowest component of ${\cal U}$ is 
\begin{equation}
{\cal O} = \tr a_i a_i^* - \tr b_i b_i^*\ ,
\end{equation}
which was identified  
in the conformal case as dual to the
Kaehler mode that introduces a small resolution of the conifold \cite{KW}. 
Again in the conformal case, $J^\mu$ is conserved, 
its dimension is exactly $3$, 
and correspondingly the dimension of the resolution 
operator ${\cal O}$ is exactly $2$.  
The fact that ${\cal O}$ is a scalar operator guarantees that its superpartner 
$J^\mu$ is an axial current (see for example Ch.~26 of \cite{Weinberg}).  
Accordingly, when the corresponding baryon number symmetry is broken, 
the Goldstone mode is a pseudo-scalar, as we have found above.

To gain some intuition for what operator the scalar
 mode corresponds to, we may substitute
an infinitesimal transformation $a_i\to (1+\lambda )a_i$,
$b_i\to (1-\lambda) b_i$ with real $\lambda$ into the action.
The potential terms are invariant, but the kinetic terms 
\begin{equation}
  S = \int d^4 x \, \tr \left[ |\partial a|^2 + 
   |\partial b|^2 \right]
\ 
\end{equation} 
are not.
For this reason we do not find a conserved Noether current.
Instead, 
\begin{equation} \label{axionop}
\delta S\sim \int d^4 x \, \lambda \tr
\left [ a_i^* \Box a_i + a_i \Box a_i^* -
b_i^* \Box b_i - b_i \Box b_i^* \right ]
\ .
\end{equation}
The dimension 4 operator that enters this action, with certain
fermionic and auxiliary field terms added,
should correspond to the scalar.

To study the saxion and axino more carefully, we should take
advantage of the full power of ${\mathcal N}=1$ supersymmetry.
The supersymmetric analog of (\ref{Jda}) is
\be \label{superJ}
\int d^4x \int d^4\theta \, {\cal U} (D^2 + \bar{D}^2) S
= -\int d^4x \int d^4 \, \theta S (D^2 + \bar{D}^2) {\mathcal U} \,,
\ee
where $S$ is a real superfield containing the axion, ${\mathcal U}$ is
defined in (\ref{Udef}), and $D^2 = D^\alpha D_\alpha$ and 
$\bar{D}^2 = \bar{D}_{\dot\alpha} \bar{D}^{\dot\alpha}$ 
are squares of the usual superderivatives that 
commute with the global supercharges.  
Current conservation means $(D^2 + \bar{D}^2) {\mathcal U} = 0$.
(In the language of \cite{Weinberg}, ${\mathcal U}$ is a linear superfield.)

We now investigate which components of 
${\mathcal U}$ act as sources for the axion and saxion.
We can write
\be
(D^2 + \bar{D}^2) S = \Phi + \Phi^\dagger \ ,
\ee
where $\Phi$ is a chiral superfield in the conventions 
of Wess and Bagger \cite{WB}:
 \eqn{PhiExpand}{
  \Phi(x) &= \phi(x) + i \theta \sigma^\mu \bar\theta \partial_\mu \phi(x)
+ \frac{1}{4} \theta\theta \bar\theta \bar\theta \Box \phi(x)  \cr
&\quad{} + \sqrt{2}\theta \psi(x) - \frac{i}{\sqrt{2}} \theta\theta \partial_\mu \psi(x)
\sigma^\mu \bar\theta + \theta\theta F(x) \ .
 }
In this language, the axion is $\Im \phi(x)$ and the saxion is $\Re \phi(x)$.
Thus, the $\theta \bar\theta$ component of $\Phi + \Phi^\dagger$ depends on
the axion while both the leading and the D-term of $\Phi + \Phi^\dagger$
depend on the saxion.

As in the bosonic case, we see that $\partial_\mu \Im \phi$ couples to $J^\mu$ through
the $\theta \bar\theta$ component of $\cal U$.  
In particular
\be
\left. {\cal U} \right|_{\theta \bar\theta} =
\theta^\alpha \bar\theta^{\dot \alpha}
\left[
i {\sigma_{\alpha \dot\alpha}}^\mu \left(\tr a_i^* \overleftrightarrow{\partial_\mu} a_i -
\tr b_i^* \overleftrightarrow{\partial_\mu} b_i \right)  + \ldots
\right] \ .
\ee
The ellipses denote bilinears in the fermions.
$J_\mu$ can be identified 
as the term inside the parentheses and is exactly what one
would expect. It is an axial current because under parity
$a_i\to a_i^*$, and $b_i\to b_i^*$.

Meanwhile, the saxion coupling
has two pieces:
$\Box \Re \phi(x)$ couples to the leading term of $\cal U$ while $\Re \phi(x)$
couples to the D-term of $\cal U$.
The D-term of $\cal U$ is easy to reconstruct from the well known
D-term of $\Phi^\dagger \Phi$:
\be
\left. (\Phi^\dagger \Phi) \right|_{\theta^2 \bar\theta^2} = \theta^2 \bar\theta^2 \left(
\frac{1}{4} \Box (\phi^* \phi) - (\partial_\mu \phi^*)(\partial^\mu \phi) + \ldots \right)  \ 
\ee
where the ellipses indicate missing fermionic and auxiliary fields.

Putting the two pieces of the saxion coupling together, we find that 
the $\Re \phi$ dependent piece of (\ref{superJ}) is
\be
\int d^4x (\Re \phi) \left(\frac{1}{2} \Box \left(\tr a_i^* a_i -\tr b_i^* b_i\right) - 
\left(\tr (\partial_\mu a_i^*)
(\partial^\mu a_i)
-\tr (\partial_\mu b_i^*) (\partial^\mu b_i) \right) + \ldots \right)
\ee
where the ellipses again denote fermionic and auxiliary field dependent terms.
This integral is in agreement with (\ref{axionop}). 

To summarize, we note that the gauge theory 
operator
corresponding to the pseudoscalar Goldstone mode is
\eqn{axionoper}{
\partial_\mu J^\mu \sim {\rm Im} 
\tr \left [ a_i^* \Box a_i -
b_i^* \Box b_i \right ] + {\rm fermion\ bilinears}\ ,
}
while the operator corresponding to the scalar is 
\eqn{saxionoper}{
\sim {\rm Re} \tr \left [ a_i^* \Box a_i -
b_i^* \Box b_i \right ] + {\rm fermion\ bilinears}
\ .}
As expected, the two operators combine into 
a natural complex operator.

\section{The Scalar Zero-Mode}
\label{Scalar}

The presence of the pseudo-scalar zero mode found in section 3, 
and the ${\cal N}=1$ supersymmetry, require the existence of a
scalar zero-mode.  In this section we argue that this zero-mode comes from
a metric perturbation that mixes with the NS-NS 2-form
potential.
Our argument is based on finding such a deformation 
(Lorentz invariant in the $t$, $x^1$, $x^2$, and $x^3$ 
directions) to first order by solving the equations of motion linearized 
around the warped deformed conifold background.  
Because we have not explicitly 
demonstrated through supersymmetry 
transformations that this mode and the pseudoscalar mode described 
in section~\ref{GENERALIZE} are part of the same multiplet of 
$d=4$, ${\cal N}=1$ supersymmetry, our argument can be regarded 
as less than airtight.  Yet we find it quite 
compelling that a previously unknown deformation 
exists with the right general properties to be the saxion.
 
The warped deformed conifold of \cite{KS} has a ${\bf Z}_2$ symmetry which
interchanges 
$(\theta_1,\phi_1)$ with $(\theta_2,\phi_2)$, i.e. it
interchanges the two $S^2$'s. 
The metric and the $F_5$ clearly have this symmetry, while
the 3-form field strengths
$F_3$ and $H_3$ actually change sign under this interchange,
so the symmetry transformations should be accompanied by a change of 
sign.\footnote{We are grateful to
E. Witten for illuminating discussions about the ${\bf Z}_2$ symmetry.}
In the gauge theory the corresponding
symmetry is the interchange of $A_1,A_2$
with $B_1,B_2$ accompanied by the charge conjugation, i.e. the interchange
of the fundamental and the antifundamental representations, in the
$SU(M+N)$ and the $SU(N)$ gauge groups.

It is easy to see that the pseudoscalar mode we found breaks this ${\bf Z}_2$
symmetry. The operator corresponding to it, (\ref{axionoper}), 
changes sign when the $A$'s and $B$'s are interchanged.
Also, from the form of the 
perturbations (\ref{NewPerturbation}) we see that 
$\delta F_3$ is even under the interchange of 
$(\theta_1,\phi_1)$ with $(\theta_2,\phi_2)$,  while $F_3$
is odd; $\delta F_5$ is odd while $F_5$ is even.

Similarly, the scalar (saxion) mode must also break the ${\bf Z}_2$
because in the field theory it breaks the symmetry between
expectation values of $|{\cal B}|$ and of $|\bar {\cal B}|$
and couples to a ${\bf Z}_2$-odd operator (\ref{saxionoper}). 
Turning on the zero-momentum
scalar modifies the background, while a zero-momentum
Goldstone mode does not.  The reason for the difference is that
the scalar changes the absolute value of  
$|{\cal B}|$ and $|\bar {\cal B}|$ while the Goldstone mode
affects only the phase.

Deformations of the warped deformed conifold that break this ${\bf Z}_2$
were considered by Papadopoulos and Tseytlin in \cite{PT}.
They wrote down the following metric ansatz:\footnote{
The radial variable $\tau$ is related to $u$ through
$d\tau =- e^{4p} du$.}
\begin{equation}
ds_{10}^2 = e^{2p-x} (e^{2A} dx_{||}^2 + du^2) + ds_5^2
\ ,
\end{equation}
where
\eqn{Arkans}{
ds_5^2 =& (e^{x+g} + a^2 e^{x-g}) \left[(e^1)^2 + (e^2)^2 \right] \cr
&+ e^{x-g} \left[ (e^3)^2 + (e^4)^2 - 2a 
\left (e^{(3} e^{1)}+ e^{(4} e^{2)}\right ) \right] + e^{-6p-x} (e^5)^2
\ ,
}
where $p,x,g,A, a$ are functions of $u$.
This metric preserves the $SO(4)$ symmetry, but the ${\bf Z}_2$
is in general broken. It is unbroken only if
$ e^g + a^2 e^{-g} = e^{-g}$, which is a relation obeyed for the KS
solution. In that case the coefficient of $(e^1)^2 + (e^2)^2$
is equal to the coefficient of $(e^3)^2 + (e^4)^2$, so that
there is symmetry between the two $S^2$'s.
On the other hand, for the
resolved conifold $a=0$ and the ${\bf Z}_2$ is broken \cite{PT,Pando-Tseytlin}.
 For the more general ansatz of \cite{PT}, which also involves a
deformation of the 2-form fields,
the system of equations was quite complicated and no new
explicit solution was found.

In this section we aim to do something simpler: 
look for the scalar zero-mode by perturbing 
around the KS solution to linear order in the ${\bf Z}_2$ breaking
perturbation.
The necessary perturbation is a mixture of the NS-NS 2-form
and the metric:
\eqn{NSPerturbation}{
  \delta B_2 &=  \chi(\tau) dg^5\ , 
\qquad \delta H_3= \chi' d\tau
\wedge dg^5 = \chi' d\tau \wedge (g^1\wedge g^4+ g^3\wedge g^2)\ , \cr
  \delta F_3 &= 0\ ,\qquad \delta F_5=0\ , \cr 
\delta G_{13} & =  \delta G_{24} = 
m(\tau)\ . 
}
To see that these components of the metric break the ${\bf Z}_2$
symmetry, we note that
\begin{equation}
(e^1)^2 + (e^2)^2 - (e^3)^2 - (e^4)^2 =
g^1 g^3 + g^3 g^1+ g^2 g^4+ g^4 g^2\ . 
\end{equation}

Now let us check consistency of this ansatz. Since
$\delta H_3 \wedge  F_3=0$,
the equation 
\begin{equation}
d\delta F_5 = \delta H_3 \wedge  F_3
\end{equation} 
is satisfied.
Since $F_5\wedge \delta H_3=0$, we must have
\begin{equation} \label{musthave}
d\delta (* F_3) = 0\ ,
\end{equation} 
where the variation comes entirely from the metric.
At linear order in $m(\tau)$ we have
\eqn{invmet}{
G^{13}& = G^{24} =- G^{11} G^{33} m(\tau)\ ,\cr
G^{11}& = G^{22}=
{2\over \epsilon^{4/3} K(\tau)\sinh^2 (\tau/2) h^{1/2}(\tau)}\ ,\cr
G^{33}& = G^{44}=
 {2\over \epsilon^{4/3} K(\tau)\cosh^2 (\tau/2) h^{1/2}(\tau)}\ .
}
Using this, we find
\begin{equation}
\delta (* F_3) = A(\tau) dx^0\wedge dx^1\wedge dx^2\wedge dx^3
\wedge d\tau \wedge dg^5\ ,
\end{equation} 
where $A(\tau)$ is a function we don't need to determine.
It follows that (\ref{musthave}) is indeed satisfied.

Now let us use the equation
\begin{equation} \label{starH}
d\delta (* H_3) = 0\ ,
\end{equation} 
which follows from the fact that variations of $F_3$ and
$F_5$ vanish. This equation is equivalent to
\begin{equation} \label{Hupper}
H^{\tau 14}= H^{\tau 32} = 0\ .
\end{equation} 
Setting $\alpha'=1$, and using
\begin{equation} 
H_{\tau 12}= {g_s M\over 2} f'\ ,
\qquad
H_{\tau 34}= {g_s M\over 2} k'\ ,
\end{equation} 
we find that (\ref{Hupper}) implies the constraint
\eqn{raiseform}{ 
\delta H_{\tau 14}=\delta H_{\tau 32}= \chi' =
{g_s M m(\tau)\over 2} (f' G^{22} + k' G^{33})\ .
} 
We will find it convenient to define
\begin{equation} \label{ztaudef}
m(\tau)= h^{1/2} K\sinh (\tau) z (\tau)=
2^{-1/3} [\sinh(2\tau) - 2\tau]^{1/3} h^{1/2} z (\tau)
\ .
\end{equation}
Then (\ref{raiseform}) becomes
\eqn{raiseformnew}{ 
\chi' =
g_s M z(\tau) [f' \coth(\tau/2)  + k' \tanh(\tau/2) ]
={1\over 2} g_s M z(\tau) {\sinh(2\tau) - 2\tau\over \sinh^2 \tau}
\ .
} 
This is equivalent to (5.21) of \cite{PT}.

Equations~\eno{starH} and~\eno{Hupper} also imply that $\delta (*_6 H_3)=0$,
where $*_6$ is calculated with the 6-d CY metric.
And with \raiseform\ in hand, one may straightforwardly show that 
$\delta(*_6 F_3) = -{1 \over g_s} \delta H_3$, where the left hand side 
is non-vanishing even though $\delta F_3 = 0$ because the definition of 
Hodge duals changes when the metric is varied.  
The upshot is that the variation we are considering preserves the 
self-duality condition of the 3-forms, $*_6 H_3 =-g_s F_3$.

\subsection{10d Einstein Equation}

It remains
to derive the equation for $m(\tau)$ by
linearizing the $13$ component of the Einstein equations.
The Einstein equation is
\eqn{lEE}{
R_{ij}& = {g_s^2\over 96} F_{iabcd}F_j^{\ abcd}+ 
{1\over 4}H_{iab} H_j^{\ ab} -{1\over 48}G_{ij} H_{abc} H^{abc}\cr
& + {g_s^2\over 4}F_{iab} F_j^{\ ab} -{g_s^2\over 48}G_{ij} F_{abc} F^{abc}
\ .
}
where we omitted the dilaton and the RR scalar terms because they do not contribute
at first order in $m(\tau)$.

We computed $\delta R_{13}$ with a computer algebra package.
The standard formula for calculating
first order perturbations to the Ricci curvature is
\[
\delta R_{ij} = \frac{1}{2} \left(
-{{\delta G_a}^a}_{;ij} - {\delta G_{ij;a}}^a + {\delta G_{ai;j}}^a + {\delta G_{aj;i}}^a
\right) \ .
\]
The covariant derivatives are taken and the index raising done with respect to the
unperturbed metric.
The first term in this expression vanishes because the metric perturbation is traceless.
However, the remaining three terms combine to give 
\begin{eqnarray}
\delta R_{13} &=& 
-\frac{3}{\epsilon^{4/3}} K^3 \sinh(\tau) z
\left[ {K'' \over K} + {1 \over 2} {h'' \over h} + {z'' \over
z}
+ {(K')^2 \over K^2} - {1 \over 2} {(h')^2 \over h^2} 
+ {K' \over K} {h' \over h} +  \right.
\nonumber \\
&&
\left. 2 {K' \over K} {z' \over z}+\coth \tau \left( {h' \over h} + 4 {K' \over K}
+ 2 {z' \over z} \right) + 2 - {1 \over \sinh(\tau)^2}
- {4 \over 9} {1 \over \sinh(\tau)^2 K^6} \right]
\nonumber \\
&=& -\frac{3}{\epsilon^{4/3}} K^3 \sinh(\tau) z
\left[
\frac{1}{2} \frac{\left( \left(K \sinh(\tau) \right)^2 (\ln h)' \right)' }{(K \sinh(\tau))^2} 
+\frac{\left( \left( K \sinh(\tau) \right)^2 z' \right)' }
{(K \sinh(\tau))^2 z} \right.
\nonumber \\
&& \left. -\frac{2}{\sinh(\tau)^2} - \frac{8}{9} \frac{1}{K^6 \sinh(\tau)^2}
+ \frac{4}{3} \frac{\cosh(\tau)}{K^3 \sinh(\tau)^2} \right] \ ,
\label{deltaR13}
\end{eqnarray}
where $z(\tau)$ is defined as in (\ref{ztaudef}).
Note that $\delta R_{13} = \delta R_{24}$ are the only nonzero first order
perturbations to the Ricci curvature.

Now we need to calculate the necessary terms linear in $m$ on the RHS 
of (\ref{lEE}).
From the 5-form we find the source
\eqn{fiveform}{ 
{g_s^2\over 96} F_{1abcd}F_3^{\ abcd} 
 =& -{g_s^2\over 4} F_{12345}^2 G^{22} G^{44} G^{31} G^{55}=
m(\tau) {g_s^2\over 4} 
F_{12345}^2 G^{11} G^{22} G^{33} G^{44} G^{55}
\cr
=& m (\tau) S(\tau)
}
where
\[
S(\tau) = {3 (g_s M)^4\over 2^{1/3} \epsilon^{20/3} h^{5/2} }
{ (\tau \coth \tau -1)^2 [\sinh (2\tau) - 2\tau]^{4/3}\over \sinh^6 (\tau)
} \ .
\]
The above can also be written as $z \tilde S$ where
\eqn{tildeS}{
\tilde S= &
{3 \over 2^{2/3} } {(g_s M)^4\over \epsilon^{20/3} h^2 }
{ (\tau \coth \tau -1)^2 [\sinh (2\tau) - 2\tau]^{5/3}\over \sinh^6 (\tau) }
\cr 
=& \frac{3}{2\cdot \epsilon^{4/3}} K^3 \sinh \tau \frac{(h')^2}{h^2} 
\ .
}

From the incompletely contracted RR and NSNS 3-form terms we get
\eqn{threeRR}{
{g_s^2\over 4}F_{iab} F_j^{\ ab} = &{g_s^2\over 2}
\left [F_{125} F_{345} G^{24} G^{55} + F_{13\tau} F_{31\tau} G^{13} G^{\tau\tau} \right]\cr
=& m(\tau){(g_s M)^2\over 8} \left[- F(1-F) G^{22} G^{44} G^{55} +  (F')^2
G^{11} G^{33} G^{\tau\tau} \right]\ ,
}
and
\eqn{threeNS}{
{1\over 4}H_{1ab} H_3^{\ ab} & = {1\over 2} \left[ H_{12\tau} H_{34\tau}
G^{24} G^{\tau\tau} +  H_{135} H_{315} G^{13} G^{55}+
\right. \cr & \left.  
H_{14\tau} H_{34\tau} G^{44} G^{\tau\tau}+
 H_{12\tau} H_{32\tau} G^{22} G^{\tau\tau} \right]\cr
=& m(\tau){(g_s M)^2\over 8}G^{55} \left[- f' k' G^{22} G^{44} 
+ (F')^2 G^{11} G^{33} 
+ (f' G^{22} + k' G^{33})^2
\right]\cr
=& m(\tau){(g_s M)^2\over 8}G^{55} \left[
(F')^2 G^{11} G^{33} 
+F(1-F) G^{22} G^{44} + 
\right. \cr & \left. 
(1-F)^2 G^{33} G^{44} +  F^2 G^{11} G^{22} \right]
\ .
}
In the last step we used the identities
$f'=(1-F)\tanh^2(\tau/2)$ and $k'=F\coth^2(\tau/2)$.
For the remaining contribution of the 3-forms, 
we use the fact that for the unperturbed solution $g_s^2 |F_3|^2 = |H_3|^2$
and only compute $|F_3|^2$:
\eqn{threeRRdiag}{
-{g_s^2\over 24} G_{13} F_{abc} F^{abc} =& -m {(g_s M)^2\over 8}
\left[ (F')^2 G^{11} G^{33} G^{\tau\tau} + {1\over 2} F^2 G^{11} G^{22} G^{55}
\right. \cr & \left.
+ {1\over 2} (1-F)^2 G^{33} G^{44} G^{55} \right]\ .
}

We see that there are some cancellations, and the contributions from
3-form sources (\ref{threeRR})-(\ref{threeRRdiag}) combine into
\eqn{threetot}{
m(\tau) {g_s^2\over 24} F_{abc} F^{abc}= & m(\tau){(g_s M)^2\over 8} G^{55} \left[
(F')^2 G^{11} G^{33} 
 + {1\over 2} (1-F)^2 G^{33} G^{44} +  
{1\over 2} F^2 G^{11} G^{22} \right]
\cr
=&
m(\tau) T(\tau)
}
where
\eqn{Ttau}{
T(\tau) = 
{3 (g_s M)^2\over 8\epsilon^4 h^{3/2}}  
{\cosh(4\tau)+ 8(1+\tau^2) \cosh(2\tau)- 24\tau \sinh(2\tau)+ 16\tau^2-9\over 
\sinh^6(\tau)}
\ .}
The above can also be written as $z \tilde T$ where
\eqn{Ttilde}{
\tilde T=& 
{3 \over 8 \cdot 2^{1/3} } 
{(g_s M)^2\over \epsilon^{4} h}
\frac{(\sinh(2\tau)  -2\tau)^{1/3}}{ \sinh^{6}(\tau)} \times \cr
&
\left(\cosh(4\tau)+ 8(1+\tau^2) \cosh(2\tau)- 
 24\tau \sinh(2\tau)+ 16\tau^2-9 \right) 
\cr
=& -\frac{3}{\epsilon^{4/3}} K^3 \sinh \tau \left[
\frac{1}{2} \frac{h''}{h} + \frac{K'}{K} \frac{h'}{h} + \coth \tau \frac{h'}{h}
\right]
\ .
}

Putting the pieces together, we can write the Einstein equation
(\ref{lEE}) as
\[
\delta R_{13} = (\tilde S + \tilde T) z(\tau) \ .
\]
Remarkably, this differential equation for $z(\tau)$ is independent of $h$.
The $\tilde S$ exactly cancels the $1/h^2$ dependent piece of
$\delta R_{13}$ and  
the $\tilde T$ precisely cancels the $1/h$ dependent terms.
The remaining differential equation for $z(\tau)$ is precisely
(\ref{deltaR13}) set to zero with $h$ set to a constant:
\eqn{zdiffeq}{
\frac{\left( \left( K \sinh(\tau) \right)^2 z' \right)' }
{(K \sinh(\tau))^2 } = \left(
\frac{2}{\sinh(\tau)^2} +\frac{8}{9} \frac{1}{K^6 \sinh(\tau)^2}
- \frac{4}{3} \frac{\cosh(\tau)}{K^3 \sinh(\tau)^2}
\right) z \ .
}

To gain confidence in this differential equation for $z(\tau)$, 
we rederived it using 
the 1d effective action of PT \cite{PT}.
In particular, 
the metric functions in (\ref{Arkans}) can be reexpressed in terms
of $z$:
\be
e^g = {1\over \cosh y - cz}, \qquad a= {\sinh y\over \cosh y - cz}
\ ,
\ee
where $c$ is a numerical constant.  With this ansatz, one can write
down a 1d effective action for $z(\tau)$, keeping only terms
of order $z^2$.  The Lagrangian is
 \eqn{ReducedL}{
  L &= \frac{\,{z(\tau)}^2\,\left( -4 + 
         6\,{K(\tau)}^3\,\sinh (\tau)\,\tanh (\tau) + 
         9\,{K(\tau)}^5\,\tanh (\tau)\,K'(\tau)
         \right) }{9\,{K(\tau)}^4}  \cr &\qquad {} - {1 \over 2}
    {K(\tau)}^2\,{\sinh (\tau)}^2\,{z'(\tau)}^2 \,.
 }
The corresponding equation of motion is precisely (\ref{zdiffeq}).  Instead of presenting a derivation of \ReducedL\ here, we have made it available online \cite{ptCheck}.

\subsection{Analysing the Scalar Mode}

The solution of (\ref{zdiffeq}) for the zero-mode
is remarkably simple:
\be \label{zresult}
z(\tau)={c_1 \coth(\tau) + c_2(\tau\coth (\tau)-1) \over
[\sinh(2\tau)-2\tau]^{1/3} }
\ ,
\ee
with $c_1$ and $c_2$ constants.
We must set $c_1=0$ in order for the zero-mode to be well behaved
at $\tau=0$.
Like the pseudoscalar perturbation, the large $\tau$ asymptotic is again 
$z\sim \tau e^{-2\tau/3}$. 
We note that the metric perturbation also has the simple form
$\delta G_{13}\sim h^{1/2} [\tau\coth (\tau)-1]
$.
Note also that the perturbed metric
$d\tilde s^6_2$ differs from the metric of the deformed
conifold by
\be
\sim (\tau\coth\tau -1) (g^1 g^3 + g^3 g^1 + g^2 g^4 + g^4 g^2)
\ ,
\ee
which grows as $\ln r$ in the asymptotic radial variable $r$.

The existence of the
scalar zero-mode makes it likely
that there is a one-parameter family of supersymmetric solutions
to the Papadopoulos-Tseytlin equations \cite{PT}, obtained
from their ansatz (\ref{Arkans}), which break the ${\bf Z}_2$
symmetry interchanging the two $S^2$'s.
We will call these conjectured backgrounds
{\bf resolved warped deformed conifolds}.
We add the word {\bf resolved} because 
both the resolution of the conifold, which is a Kaehler deformation,
and these resolved warped deformed conifolds
break the ${\bf Z}_2$ symmetry.
(Note that at the special point where the ${\bf Z}_2$ breaking 
parameter vanishes, the resolved warped deformed conifold becomes an
ordinary warped deformed conifold.)

As we explained in section 4, in the dual gauge theory
turning on the ${\bf Z}_2$ breaking corresponds to
the transformation
${\cal B} \to \xi {\cal B}$,
$\bar{\cal B} \to \xi^{-1} \bar{\cal B}$ on the baryonic branch.
Therefore, $\xi$ is dual to the ${\bf Z_2}$ breaking parameter of the 
warped deformed conifold.

One might ask whether our resolved warped deformed conifolds
are 
still of the form 
$h^{-1/2} dx_{||}^2 + h^{1/2} d\tilde s_6^2$
where $d\tilde s_6^2$ is Ricci flat.  
At linear 
order in our perturbation, our conifold
metric $d\tilde s_6^2$ is indeed Ricci flat: the first order corrections
vanish if (\ref{zdiffeq}) is satisfied.
We also showed (see below \raiseformnew) that the complex 3-form field
strength $G_3= F_3- {i\over g_s} H_3$ remains
imaginary self-dual at linear order,
i.e.~$*_6 G_3 =i G_3$.
It will be interesting to see if these properties continue to hold for the
exact solution.

\section{Compactification and Higgs Mechanism}

As we argued above, the non-compact warped deformed conifold
exhibits a supergravity
dual of the Goldstone mechanism. It was crucial for our arguments
that the $U(1)_B$ symmetry is not gauged in the field theory, and   
the appearance of the Goldstone boson in the supergravity dual confirms
the symmetry is global.

If the warped deformed conifold is embedded into a flux compactification
of type IIB string on a 6-dimensional CY manifold, then we expect
the global $U(1)_B$ symmetry to become gauged, because the square
of the gauge coupling becomes finite. 
In the compact case we may write $\delta C_4 \sim
\omega_3 \wedge A$, where $\omega_3$ is harmonic in the full compact
case and $A$ is the 4-d gauge field. If we ignore subtleties with the self-duality
of the 5-form field strength, then the kinetic terms for it is
\be
{1\over 2 g_s^2} \int d^{10} x \sqrt{-g} F_5^2\ .
\ee
Substituting $F_5=F_2\wedge \omega_3$ and reducing to 4 dimensions, we
find the $U(1)$ kinetic term
\be
{1\over 2 g^2} \int d^4 x F_2^2\ ,
\ee
where
\be
{1\over g^2} \sim {1\over g_s^2} \tau_m\ ,
\ee
where we assumed that the effect of compactification is to introduce a cut-off
at $\tau_m\gg 1$.

The finiteness of the gauge coupling in the compact case means that
the Goldstone mechanism should turn into a Higgs mechanism.
The Goldstone boson $a$ enters as a gauge parameter of  
$A$ and gets absorbed by the $U(1)$ gauge field to make a
massive vector field. As usual in the supersymmetric Higgs mechanism,
the scalar acquires the same mass. As a result, we find an
${\cal N}=1$ massive vector supermultiplet containing
a massive vector, a scalar (the Higgs boson), and their fermion   
superpartners.

While in the non-compact case D-strings are global strings,
in the compact case they should be interpreted as
Abrikosov-Nielsen-Olesen vortices of an Abelian-Higgs
model, where the charged chiral superfields
breaking the gauge symmetry are the baryon operators ${\cal B}$  
and $\bar {\cal B}$.
Representation of D-strings by ANO vortices
in low-energy supergravity was recently advocated in \cite{Dvali},
where it was
argued that such vortices can be BPS saturated if
an all-important Fayet-Iliopoulos D-term is included.
Note, however, that D1-branes placed at the bottom of the throat
embedded into a flux compactification cannot be BPS. Since there is
a finite number $K$ of NS-NS
flux units through a cycle dual to the 3-sphere \cite{GKP},
 the D-string charge
takes values in ${\bf Z}_K$: now $K$ D-strings can break
on a wrapped D3-brane \cite{Copeland}. Correspondingly, we
do not expect the ANO vortex duals to be BPS saturated.
We hope to return to investigation of these issues
in future work.

\section{A double T-duality on the KS solution}

In this section we find the supergravity background describing
D-strings at $\tau=0$ that are smeared over their transverse directions
$x^2, x^3$ as well as over the $S^3$. The resulting solution has $B_{23}$
turned on, and may be thought of as a non-commutative deformation of the
warped deformed conifold solution. A strategy that was useful in obtaining
the non-commutative generalization of $AdS_5\times S^5$ was to use
double T-duality along the $x^2, x^3$ directions 
\cite{Hashimoto:1999ut,Maldacena:1999mh}.

We will use the rules for T-duality from page 30 of the paper of
Bergshoeff, Hull, and Ortin (BHO) \cite{Bergshoeff:1995as}. 
We want to T-dualize the KS supergravity solution first in the $z$ direction
and then in the $z \cos \theta  - y \sin \theta$ direction where our
gauge theory coordinates are $x$, $y$, $z$, and $t$.
The BHO rules simplify quite a bit because to start with there is no
NS-NS $B_2$ or RR $C_2$ potential in the $z$ direction.  Moreover,
the metric is diagonal, and the axion field is turned off.

Let the hatted fields be the result of the two T-dualities.  Two
new coordinates are needed at the end of the day:
\eqn{Tdualcoord}{
{y}' = y \cos \theta  + z \sin \theta
}
and the T-dual $z'$ of 
$z \cos \theta  - y \sin \theta$.  
Recall that the BHO rules are worked out in the string frame.

The transformation rules for the dilaton, metric, and NS-NS $B$ field
can be found in many places in addition to the BHO paper.
After two T-dualities, the string frame metric becomes
\eqn{tendmetric}{
ds_{10}^2 = h(\tau)^{-1/2} (-dt^2 + dx^2 + \chi(\tau)(dy^2+dz^2))
+h(\tau)^{1/2} ds_6^2 \ ,
}
where $ds_6^2$
is the usual warped deformed conifold metric (\ref{conifoldmetric})
and
\eqn{nonextfactor}{
\chi(\tau)^{-1} = \cos^2 \theta + h(\tau)^{-1} \sin^2 \theta \ 
}
can be written in terms of the usual warp factor $h(\tau)$.
The dilaton becomes
\eqn{dilaton}{
e^{2\hat \phi} = g_s^2 \chi(\tau) \ .
}
(We have departed from the convention of the previous sections where
$e^\phi$ was independent of $g_s$.)
The NS-NS two form in the conifold directions is unchanged:
\eqn{angB}{
\hat B_{\alpha \beta} = B_{\alpha \beta} \ .
}
At the same time, $\hat B_2$ picks up a $y'z'$ dependent
piece:
\eqn{fourdB}{
\hat B_{y'z'} = \tan \theta (1-h^{-1} \chi) \ .
}
This $y'z'$ dependent piece becomes more familiar
if we add a constant $\tan \theta$:
\eqn{Balt}{
\hat B_{y'z'} - \tan \theta =  - h^{-1} \chi \tan \theta
}
which matches (2.1) of the Maldacena-Russo paper 
\cite{Maldacena:1999mh} up to a sign.
Our sign conventions appear to be necessary
to cancel the minus sign in $F_5 = dC_4 - C_2 \wedge H_3$.

T-duality transformation rules for the RR potentials are less well known, and 
we really rely on the BHO paper here.  Unfortunately, our 
conventions for the SUGRA fields are rather different from the BHO paper.
We have fixed signs and overall normalizations by checking these
transformation rules using the SUGRA 
equations of motion.
For the four form and two form potentials in the gauge theory directions,
\eqn{fourdC4}{
\hat C_{txy'z'} =  \chi C_{txyz}  \cos \theta \ ,
}
and
\eqn{fourdC2}{
\hat C_{tx} = C_{txyz} \sin \theta 
}
which agrees with (2.1) of the Maldacena-Russo paper \cite{Maldacena:1999mh}.  
To
show agreement, we have used the
fact that in these types of warped compactifications,
$g_s C_{txyz} = h^{-1}$ where $h$ is the warp factor and also
that $F_5 = dC_4 - C_2 \wedge H_3$.
As in \cite{Maldacena:1999mh}, we find
\eqn{RRmatch}{
g_s \hat C_{tx} = h^{-1} \sin \theta \ ,
\qquad 
g_s \hat F_{tx y' z' \tau} = \chi \partial_\tau h^{-1} \cos \theta \ .
}

In the KS solution, we also have to trace how $F_3$
transforms:
\eqn{angC4}{
\hat C_{\alpha \beta y' z'} = -C_{\alpha \beta} h^{-1} \chi \sin \theta \ ,
}
\eqn{angC2}{
\hat C_{\alpha \beta} =  C_{\alpha \beta}  \cos \theta \ .
}
These rules are reminiscent of the results (2.9--2.12) 
of Mateos, Pons, and Talavera \cite{Mateos:2002rx}.  
However, their T-duality prescription
is not completely the same as ours.  The minus sign
in $\hat C_{\alpha \beta y' z'}$ is necessary to cancel a
$C_2$ dependent piece in $F_5$.
 From the equation for $\hat C_{\alpha \beta y' z'}$, we have
\eqn{RRmatchmore}{
\hat F_{tx\alpha\beta\gamma}=-\hat C_{tx} H_{\alpha\beta\gamma}\ ,
}
and the component 
\eqn{Fdual}{
\hat F_{y' z'\alpha\beta\gamma} = -h^{-1} \chi \sin \theta F_{\alpha \beta \gamma} 
}
related to it by
Hodge duality.

We have checked that the hatted fields satisfy the string frame
dilaton equation of motion
\eqn{stringeomdil}{
d * d e^{-2 \hat \phi} = e^{-2 \hat \phi} 
\hat H_3 \wedge * \hat H_3 - \hat F_3 \wedge * \hat F_3
}
and the equation of motion for $\hat F_3$:
\eqn{stringeomF3}{
d * \hat F_3 = \hat F_5 \wedge \hat H_3 \ .
}

\section{Discussion}

Our paper sheds new light on the physics of the cascading 
$SU(N+M)\times SU(N)$ gauge theory, whose supergravity dual is the warped
deformed conifold \cite{KS}. In the infrared the theory is not 
in the same universality class as the pure glue ${\cal N}=1$
supersymmetric $SU(M)$ theory: the cascading theory contains
a massless pseudoscalar glueball, as well as global strings dual to
the D-strings placed at $\tau=0$ in the supergravity
background. 

As suggested in \cite{KS,Aharony} and reviewed in section 4 above, the infrared field theory 
is better thought of as $SU(2M)\times SU(M)$ on the baryonic branch, i.e.
with baryon operators (\ref{baryonops}) having expectation values.
Since the global baryon number symmetry, $U(1)_B$, 
is broken by these expectation values, 
the spectrum must contain Goldstone bosons.
However, these modes were not found in the spectrum of
supergravity fluctuations until now. In this paper we find explicitly the
pseudo-scalar Goldstone mode, and we construct at linear order a Lorentz-invariant deformation of the background which we argue is a zero-momentum state of the scalar superpartner of the Goldstone mode.  These calculations confirm the validity 
of the baryonic branch interpretation of the gauge theory.
This resolves a puzzle about the dual of the D-strings
at $\tau=0$: they are the solitonic strings that couple to these
massless glueballs. We further argue that, upon embedding
this theory in a warped compactification, the global
$U(1)_B$ symmetry becomes gauged; then the gauge symmetry
is broken by the baryon expectation values through a
supersymmetric version of the
Higgs mechanism. Thus, in a flux compactification,
we expect the D-string to be dual to an Abrikosov-Nielsen-Olesen
vortex.  

In \cite{KS} it was argued that there is a limit, $g_s M\to 0$,\footnote{
No string theoretic description of this limit is yet available, because it
is the opposite of the limit of large $g_s M$ where the supergravity
dscription is valid. }
where the physics of the cascading gauge theory should approach that of the
pure glue ${\cal N}=1$ supersymmetric $SU(M)$ gauge theory.
The reader may wonder
how this statement can be consistent with the presence
of the Goldstone bosons. We believe that it can. 
Returning to the $SU(2M)\times SU(M)$ gauge theory discussed
in section 4, we expect that
in the limit $g_s M\to 0$ the scale $\Lambda_{2M}$ of the
$SU(2M)$, i.e. that of the baryon condensates,
is much higher than the scale $\Lambda_M$ of the $SU(M)$. 
Hence, the decay constant $f_a$ should be much greater
than the confinement scale $\Lambda_M$. Since the
Goldstone boson interactions at the confinement
scale are suppressed by powers of $\Lambda_M/f_a$, 
they appear to decouple from the massive glueballs containing
the physics of the pure glue supersymmetric $SU(M)$ gauge theory.
Obviously, this heuristic argument needs to be subjected
to various checks.

Our work opens new directions for future research. 
Turning on finite scalar perturbations is expected to give rise to
a new class of Lorentz invariant supersymmetric backgrounds,
{\bf the resolved warped deformed conifolds}, which preserve the $SO(4)$
global symmetry but break the ${\bf Z}_2$ symmetry of the
warped deformed conifold.
The ansatz for such backgrounds was proposed in
\cite{PT}. 
We have argued that these conjectured backgrounds are dual to the
cascading gauge theory on the baryonic branch. 
It would be desirable to find them explicitly,
and to confirm their supersymmetry.

A more explicit construction of the global string as
a soliton in the gauge theory is desirable. 
It is also interesting to explore the 
consequences of our results for cosmological modeling.

We hope that future work leads
to new insights into the remarkable physics of cascading
gauge theories.

\section*{Acknowledgments}
We are grateful to D. Berenstein, J. Maldacena,
J. Polchinski, A. Polyakov, R. Roiban, N. Seiberg and especially 
O. Aharony, M. Strassler, and E. Witten for useful
discussions.  
We would like to thank M.~Srednicki for telling us about the saxion.
The work of SSG~was supported in part by the Department of 
Energy under Grant No.\ DE-FG02-91ER40671, and by the Sloan Foundation.
The work of CPH was supported in part by
the National Science Foundation Grant No. PHY99-07949.
  The work of IRK was supported in part by
the National Science Foundation Grants No.
PHY-0243680 and PHY-0140311.
Any opinions, findings, and conclusions or recommendations expressed in
this material are those of the authors and do not necessarily reflect
the views of the National Science Foundation.

\appendix

\section{Formulae for the KS solution}

Here we collect some necessary formulae from \cite{KS} (for
reviews see \cite{Herzog}).

The ten dimensional metric for the KS solution  is
\be \label{10dmetric}
ds_{10}^2 = h(\tau)^{-1/2} (-dt^2 + dx^2 + dy^2+dz^2)
+h(\tau)^{1/2} ds_6^2 \ ,
\ee
where
\eqn{conifoldmetric}{
ds_6^2 = {\epsilon^{4/3} K(\tau) \over 2}
\Bigg[ {1 \over 3K^3} (d\tau^2 + (g_5)^2) 
} \[ 
 + \cosh^2 \left ({\tau\over 2}\right ) ((g^3)^2 + (g^4)^2)
+ \sinh^2 \left ({\tau\over 2}\right ) ((g^1)^2 + (g^2)^2) \Bigg] \ 
\]
is the usual warped deformed conifold metric.
The one forms are given in terms of angular coordinates as
\begin{eqnarray} \label{gbasis}
g^1 = {e^1-e^3\over\sqrt 2}\ ,\qquad
g^2 = {e^2-e^4\over\sqrt 2}\ , \nonumber \\
g^3 = {e^1+e^3\over\sqrt 2}\ ,\qquad
g^4 = {e^2+ e^4\over\sqrt 2}\ , \nonumber \\
g^5 = e^5\ ,
\end{eqnarray}
where
\begin{eqnarray} \label{ebasis}
e^1\equiv - \sin\theta_1 d\phi_1 \ ,\qquad
e^2\equiv d\theta_1\ , \nonumber \\
e^3\equiv \cos\psi\sin\theta_2 d\phi_2-\sin\psi d\theta_2\ , \nonumber\\
e^4\equiv \sin\psi\sin\theta_2 d\phi_2+\cos\psi d\theta_2\ , \nonumber \\
e^5\equiv d\psi + \cos\theta_1 d\phi_1+ \cos\theta_2 d\phi_2 \ .
\end{eqnarray}
Note that 
\be \label{Keqn}
K(\tau)= { (\sinh (2\tau) - 2\tau)^{1/3}\over 2^{1/3} \sinh \tau}
\ .
\ee
The warp factor is
\be \label{intsol}
h(\tau) = (g_s M\alpha')^2 2^{2/3} \varepsilon^{-8/3} I(\tau)\ ,
\ee
where
\be \label{Itau}
I(\tau) \equiv
\int_\tau^\infty d x {x\coth x-1\over \sinh^2 x} (\sinh (2x) - 2x)^{1/3}
\ .
\ee

The NS-NS two form field is
\be \label{B2}
B_2 = {g_s M \alpha'\over 2} [f(\tau) g^1\wedge g^2
+  k(\tau) g^3\wedge g^4 ]\ ,
\ee
\begin{eqnarray}
H_3 = dB_2 &=& {g_s M \alpha'\over 2} \bigg[
d\tau\wedge (f' g^1\wedge g^2
+  k' g^3\wedge g^4)
\nonumber \\
&& \left. + {1\over 2} (k-f)
g^5\wedge (g^1\wedge g^3 + g^2\wedge g^4) \right]\ ,
\end{eqnarray}
while the RR three form field strength is
\begin{eqnarray}
F_3 = {M\alpha'\over 2} \left \{g^5\wedge g^3\wedge g^4 + d [ F(\tau)
(g^1\wedge g^3 + g^2\wedge g^4)]\right \} \nonumber \\
= {M\alpha'\over 2} \left \{g^5\wedge g^3\wedge g^4 (1- F)
+ g^5\wedge g^1\wedge g^2 F + F' d\tau\wedge
(g^1\wedge g^3 + g^2\wedge g^4) \right \}\ .  
\end{eqnarray}
The auxiliary functions in these forms are
\begin{eqnarray}
F(\tau) &=& {\sinh \tau -\tau\over 2\sinh\tau}\ ,
\nonumber \\
f(\tau) &=& {\tau\coth\tau - 1\over 2\sinh\tau}(\cosh\tau-1) \ ,
\nonumber \\
k(\tau) &=& {\tau\coth\tau - 1\over 2\sinh\tau}(\cosh\tau+1)
\ .
\end{eqnarray}

Two harmonic forms important in section 2 are
\be \label{defo2}
\omega_2 = \frac{1}{2}(g^1 \wedge g^2 + g^3 \wedge g^4) = \frac{1}{2}
(\sin\theta_1 d\theta_1 \wedge d\phi_1 - \sin \theta_2 d\theta_2 \wedge d\phi_2) \ ,
\ee
\be \label{defo3}
\omega_3 = g^5 \wedge \omega_2 \ .
\ee

\section{The D-string: Blow up or dud?}

In \cite{Herzog:2001fq} 
it was shown that, under the influence of RR-flux through
the $S^3$, F-strings blow up into a D3-brane wrapped over an $S^2$
at a finite azimuthal angle within the $S^3$ at $\tau=0$.
Here we explore the possibility that the NS-NS field within
the $R^3$ fiber makes
D1-branes blow up into an $S^2$ at some small
$\tau$. 

We wrap the D3-brane in such a way that the pull back of $g^1$ and $g^2$
is non-vanishing, while that of $g^3$, $g^4$, and $g^5$ vanishes.
In particular, we choose 
standard angular coordinates $(\theta, \phi)$
on an $S^2$ such that $\phi = \phi_1 = -\phi_2$, $\theta=\theta_1 = -\theta_2$, and
$\psi=0$.  For this choice
it follows from (\ref{gbasis}) and (\ref{ebasis}) that  
the pull back $g^1|_{S^2} = -\sqrt{2} \sin \theta d\phi$ and $g^2|_{S^2} = \sqrt{2} d\theta$.

The tension of the wrapped D3-brane is
\be \label{T3}
 L = T_3 \sqrt{ \det (G+ B_2- (2\pi \alpha') F_2) } 
\ee
The gauge field 
$$F_2 = \frac{q}{2} \sin \theta d\theta \wedge d\phi$$ 
is quantized, 
$$\int F_2 = 2\pi q$$
where $q$ is the number of D1-branes.
The NS-NS two form (\ref{B2}) restricted to the two sphere becomes
\begin{equation}
B_2|_{S^2} = g_s M \alpha' f(\tau) \sin\theta d\theta \wedge d\phi \ .
\end{equation}
The induced metric $G$ on the D3-brane can be calculated
from (\ref{10dmetric}) and is
\be
ds_{ind}^2 = h(\tau)^{-1/2} (-dt^2 + dx^2 ) + h(\tau)^{1/2} \epsilon^{4/3} K \sinh(\tau/2)^2
(d\theta^2 + \sin^2 \theta d\phi^2) \ .
\ee

Substituting the metric, $B_2$, and $F_2$ into the tension, 
we find
that (\ref{T3}) becomes
\be \label{fulltension}
L = T_3 \epsilon^{4/3} \left[ K^2 \sinh(\tau/2)^4 +  \left( f - \frac{x}{2} \right)^2 2^{-2/3} I(\tau)^{-1}
\right]^{1/2} \ ,
\ee
where $I(\tau)$ is given by (\ref{Itau}) and $x = 2\pi q/g_s M$.

We want to analyze this tension to see if there is a minimum for small
but non-vanishing $\tau$.  As a first step, we expand $I(\tau)$ in a power
series near $\tau=0$:
$$ I(\tau) =  a_0 - {6^{-1/3}\over 3} \tau^2  + {\mathcal O}(\tau^4) 
\ ,
$$
and $a_0= 0.71805\ldots$.
We assume that $x$ is also small and
write $L$ as a power series in $x$ and $\tau$:
\be
L = T_3 \epsilon^{4/3} 2^{-1/3} \left[
\frac{x^2}{4a_0} + \frac{x^2 \tau^2}{12 \cdot 6^{1/3} a_0^2} - 
\frac{x \tau^3}{12 a_0} + \frac{\tau^4}{4 \cdot 6^{2/3}} + \ldots \right]^{1/2} \ .
\ee
Clearly, $\tau=0$ is a minimum of this function.

To minimize $L$ at small but nonzero 
$\tau$ and $x$, we would need to solve the following
quadratic equation
\be
\frac{1}{\tau} \frac{d L}{d \tau} \sim \left(\frac{6^{1/3} a_0 \tau}{x}\right)^2 
- \frac{3}{2} \left(\frac{6^{1/3} a_0 \tau}{x}\right) + 1 = 0 \ .
\ee
The discriminant of this quadratic equation is
negative and there is no minimum for real and nonzero
$\tau$; the only minimum is at $\tau = 0$.

Plotting the function (\ref{fulltension}) for a finite $x$, we see that, again, the only
minimum is at $\tau =0$. Therefore, the blow-up of D1-branes into
a wrapped D3-brane does not occur. This should be contrasted with the
fundamental strings which do blow up into a D3-brane wrapped over an $S^2$
within the $S^3$ at $\tau=0$ \cite{Herzog:2001fq}. As a result of this
effect, the fundamental string charge takes values in ${\bf Z}_M$, and the
fundamental strings are not BPS saturated objects.

The D-string charges take values in ${\bf Z}$. Also, a D-string at $\tau=0$
is the object with the smallest tension for a given charge.
These facts suggest that the D-strings at $\tau=0$ are
BPS saturated.
Consider, however, the T-dual of the 
conifold with $M$ D5-branes wrapped over the $S^2$ at the tip.
This is provided by the type IIA theory with an NS5-brane spanning the $12345$
directions, separated from an NS5-brane spanning
the $12367$ directions by some distance along the compact $x^9$ direction.
Under the T-duality, the $M$ wrapped D5-branes turn into $M$ D4-branes
spanning the $1239$ directions and suspended between the NS5-branes
\cite{Dasgupta}.
A D-string of type IIB theory turns into a D2-brane of type IIA
wrapped over the compact direction, spanning for example the $19$
directions. The fact that this object is parallel
to the D4-branes appears to break the supersymmetry.

Upon lifting to M-theory the NS5-branes and D4-branes deform and
merge into an M5-brane \cite{HOO,ewmqcd,BIKSY}.
This M-theory configuration has two compact coordinates,
$x^9$ and $x^{10}$. An M2-brane wrapping the $x^{10}$ direction is the dual
of the F-string on the conifold, while 
an M2-brane wrapping the $x^{9}$ direction is the dual of the D-string.
 From this point of view it is less clear what causes 
the violation of the BPS condition.
In any case, it would be interesting to determine whether 
the D-strings at $\tau=0$ are
BPS through a direct calculation with the warped deformed 
conifold background of \cite{KS}.

\end{document}